\documentclass{pasj00}
\usepackage{psfig}
\newcommand{\msol}{\mbox{M$_\odot$}}
\newcommand{\dfig}[2]            % Double FIGures (put two figures side by
{
  \begin{center}
    \begin{minipage}[t]{0.45\textwidth}
        \psfig{file=#1,height=0.75\textwidth,angle=270}
    \end{minipage}
    \hfill
    \begin{minipage}[t]{0.45\textwidth}
        \psfig{file=#2,height=0.75\textwidth,angle=270}
    \end{minipage}
  \end{center}
}

\newcommand{\dfigs}[2]           % Double FIGures (put two figures side by
{
  \begin{center}
    \begin{minipage}[t]{0.45\textwidth}
        \psfig{file=#1,height=0.6\textwidth,angle=270}
    \end{minipage}
    \hfill
    \begin{minipage}[t]{0.45\textwidth}
        \psfig{file=#2,height=0.6\textwidth,angle=270}
    \end{minipage}
  \end{center}
}

\newcommand{\sfig}[1]            % Single FIGure (put one figure of the same
{
  \begin{center}
    \begin{minipage}[t]{0.45\textwidth}
        \psfig{file=#1,height=1.2\textwidth,angle=270}
    \end{minipage}
  \end{center}
}

\begin{document}
\title{Milli-arcsecond--scale Spectral Properties and Jet Motions in M87}
\SetRunningHead{Dodson et al.}{Spectral Properties and Jet Motions in M87}
\Received{2005/07/08}%{yyyy/mm/dd}
\Accepted{2005/11/11}%{yyyy/mm/dd}

\author{Richard \textsc{Dodson},
%\thanks{Present Address, OAN, Espa\~na. r.dodson@oan.es}, 
Philip G. \textsc{Edwards}, 
Hisashi \textsc{Hirabayashi}} %
\affil{The Institute of Space and Astronautical Science,
Japan Aerospace Exploration Agency, \\
3-1-1 Yoshinodai, Sagamihara, Kanagawa 229-8510}


\maketitle

\begin{abstract}
We have combined high resolution VLBI Space Observatory Programme
(VSOP) data at 1.6 and 4.8 GHz with Very Long Baseline Array (VLBA)
data at higher frequencies and with similar resolutions to study the
spectral properties of the core of M87 with milliarcsecond resolution.
The VSOP data allow a more accurate measurement of the turn-over
frequency, and hence more reliable determination of associated
physical parameters of the source.  Comparison of the images with
previously published images yields no evidence for significant motion
of components in the parsec-scale jet. In addition, the brightness
temperatures obtained from model-fits to the core are well below the
inverse Compton limit, suggesting the radio emission we are observing
is not strongly Doppler boosted.

\end{abstract}

%\maketitle
\section{Introduction}

The identification of the radio source Virgo~A with the giant
elliptical galaxy M87 was one of the first such associations to be
made and demonstrated the gains in angular resolution obtainable from
interferometric observations \citep{bol49}.  M87 (NGC4486, 3C\,274,
J1230+1223) lies near the center of the Virgo Cluster and 
was already well known due to its ``curious straight ray''
(\cite{cur18}). This jet extends over 20$''$ at optical (e.g.,
\cite{spa96}) and X-ray (e.g., \cite{mar02}) energies and has been
well studied at radio wavelengths (e.g., \cite{owe89}). VLA
observations at 90\,cm have revealed the jet extends considerably
further, appearing to flow into two large bubbles in the intergalactic
medium (\cite{owe00}).

Hubble Space Telescope (HST) spectroscopic observations indicate the
presence of supermassive black hole in the core of the galaxy, with $3
\times 10^9$ \msol\ contained within the central 3\,pc (\cite{har94})
and the process of accretion of material towards the core is believed
to power the jet.  At a distance of only 16.75\,Mpc (\cite{whi95}),
high angular resolution studies correspond to the highest linear
resolutions obtainable for AGN, with 1\,mas corresponding to 0.08\,pc.
The inner parsec-scale jet has been studied at increasingly higher
radio frequencies, revealing that jet collimation begins within
several tens of Schwarzschild radii of the core but is not complete
until several thousand Schwarzschild radii
(\cite{jun99,ly04a,krich_04, krich_05}).  A comparison of Chandra and
HST data indicates that the X-ray flux dominates closer to the core
but decreases relative to the optical flux with increasing core
distance (\cite{mar02}).  A detection of M87 has also been reported at
TeV energies from a region which includes the core and inner jet,
although systematic pointing errors prevented the identification of
the gamma-ray production site with any feature at other wavelengths
(\cite{aha03}).  The TeV detection has, however, yet to be confirmed
(\cite{leb04}).

The study of motions of features in the jet has yielded a number of
surprises. Early VLBI observations revealed that apparently
superluminal motions (i.e., $v_{\rm app} > c$) were common in
radio-loud active galactic nuclei (AGN), however the first multi-epoch
1.6\,GHz VLBI observations of M87 suggested that the parsec-scale jet
displayed sub-luminal (i.e., $v_{\rm app} < c$), with an apparent
speed of 1.1$\pm$0.3\,mas\,yr$^{-1}$, corresponding to
0.28$\pm$0.08\,$c$ (\cite{rei89}).  Observations with the VLA at
15\,GHz at six epochs spanning 9 years resulted in the detection of
apparent motions between 0.5\,$c$ and $\sim c$ for the bright knots
over 1\,kpc from the core, but apparent speeds of $\sim 2 c$ for two
features $\sim 200$\,pc from the core (\cite{bir95}).  Further VLBI
observations at 22\,GHz yielded only upper limits, with apparent
speeds of $\lesssim 0.3 c$ for components within 4\,mas of the core
(\cite{jun95}) and the analysis of further 1.6\,GHz VLBI images
suggested features out to 160\,mas were stationary ($-0.03\pm0.02 c$,
\cite{bj95}).

In contrast, a series of five HST images resulted in the measurement
of apparent speeds of 2.6\,$c$--6\,$c$ for components between 0.87$''$
and 6$''$ of the core (\cite{bir99}).  An apparent speed of
0.63$\pm$0.27\,$c$ was measured for the innermost resolvable
component, 160\,mas from the core (\cite{bir99}).  More recently,
\citet{kel04} have reported an apparent speed of
0.14$\pm$0.07\,mas\,yr$^{-1}$, or 0.04$\pm$0.02\,$c$ for a component
6\,mas from the core, based on 9 epochs of 15\,GHz observations over
the six years from 1995 to 2001. A component detected between 1 and
2\,mas from the core showed no net motion over the same time range.
Finally, at 43\,GHz \citet{ly04b} have reported lower limits on the
apparent speed of 0.25\,$c$ to 0.40\,$c$ for the jet, and 0.17\,$c$
for the counter-jet.

We report here on contemporaneous observations of M87 made as part of the
VLBI Space Observatory Programme (VSOP; \cite{hir98,hir00}) and the
Very Long Baseline Array (VLBA).  The multi-frequency data allows the
evolution of the spectral index with core distance to be studied, and
comparison with previously published images enables component motions
to be examined.

\section{The Observations}

A series of VSOP observations of M87 have been made at 1.6\,GHz and
4.8\,GHz over the lifetime of the mission (see, e.g.,
\cite{jun00,bir02}).  We have selected from the VSOP data archive 1.6
and 4.8\,GHz observations made over a three day period in March 2000,
and combined them with VLBA data from observations made three weeks
later, to minimize the effects of source variability.  Details of the
observations presented here are given in table~1.

\begin{table*}[htbp] 
\begin{center} 
\caption{Observing Sessions}\label{Observation log} 
\begin{tabular}{cccl} 
\hline \hline
Observation code & Frequency & Date & Ground Telescopes\footnotemark[$*$]  \\
                 &  [GHz]    &      &                \\ 
\hline
w022a7 & ~1.646 & 2000 Mar 22 & HALCA, VLBA, Ti, Ro \\ 
w040a5 & ~4.812 & 2000 Mar 20 & HALCA, VLBA \\ 
bj031a & 15.365 & 2000 Apr 07 & VLBA  \\ 
bj031a & 22.233 & 2000 Apr 07 & VLBA  \\
bj031b & 43.217 & 2000 Apr 09 & VLA, VLBA-Hn  \\ \hline
 &  &  & \\ 
\multicolumn{4}{@{}l@{}}{\hbox to 0pt{\parbox{180mm}{\footnotesize
\footnotemark[$*$] Ground telescopes: 
Hn = Hancock 25\,m; 
Ti = Tidbinbilla 70\,m; 
Ro = Robledo 70\,m; \\
VLA = phased VLA;
VLBA = full VLBA.\\
}\hss}}
\end{tabular}
\end{center}
\end{table*}

\begin{table*}[htbp] 
\begin{center} 
\caption{Observation Parameters}
\begin{tabular}{ccccc} 
\hline \hline
            & Resolution  & Residual & Residual noise &  Core Brightness  \\
 Frequency  & (beam size) & noise level & for 1.23$\times$0.96 beam &  Temperature \\

  [GHz]    &   [mas]    & [mJy/beam] & [mJy/beam] & [K/$10^{10}$] \\ 

\hline
 ~1.646 & $1.2\times1.0$ & 1.1 & 1.1 & 2.8 \\
 ~4.812 & $0.7\times0.5$ & 0.2 & 0.3 & 3.2 \\
 15.365 & $0.9\times0.5$ & 2.0 & 2 &   1.8 \\
 22.233 & $0.6\times0.3$ & 1.5 & 3 &   1.4 \\
 43.217 & $0.3\times0.2$ & 1.5 & 5 &   5.7 \\
\hline
\end{tabular}
\end{center}
\end{table*}

The VSOP observations were made in the standard VSOP observing mode,
with two 16\,MHz bandwidth channels of two-bit sampled, left circular
polarization (LCP) data recorded at all stations (\cite{hir00}).
HALCA data were received and recorded at the Tidbinbilla, Goldstone,
Robledo and Green Bank tracking stations.  The ten telescopes of the
Very Long Baseline Array (VLBA) co-observed at both 1.6 and\,4.8 GHz,
and in addition the Tidbinbilla and Robledo 70\,m telescopes
participated in the 1.6\,GHz observation.

The VLBA observations on 2000 April 7 recorded 8 IFs (both
polarisations) of 8\,MHz bandwidth alternating between 15 and
22\,GHz approximately every twenty minutes. North Liberty was found to
have poor signal to noise and was excluded.  The 43\,GHz observation
on 2000 April 9 was made with 8 IFs of 8\,MHz bandwidth, and included
the phased VLA, but not Hancock.

\section{Data Reduction}

The data were reduced in AIPS\footnote{Astronomical Image Processing
Software developed and maintained by the National Radio Astronomy
Observatory.} using the standard VSOP and VLBA methods, and the VLBA
pipeline scripts where possible. The VLBA antenna system temperatures
are recorded within the data files. The nominal system temperatures
for the respective observing frequencies were used for HALCA
(\cite{hir00}) and also DSN telescopes, and the VLA antenna
temperatures were scaled to 3C286 (also observed in the
experiment). All of these were applied to the data via the task {\sc
antab}. After this initial amplitude calibration, delay and phase
calibration was undertaken with {\sc fring}, before exporting the data
for deconvolution and imaging in Difmap (\cite{she97}). For further
image analysis we used MIRIAD (\cite{sau95}) for alignment,
smoothing and spectral index calculation. Finally we used Octave
\citep{octave} to calculate the turn-over frequency using the
minimisation routines therein.

\section{Images}

\begin{figure}
%% \dfig{m87_1.6.ps}{m87_4.8.ps}
%% \dfig{m87_15.ps}{m87_22.ps}
%% \sfig{m87_43.ps}
\begin{center}
%  \begin{minipage}[t]{0.45\textwidth}
      \psfig{file=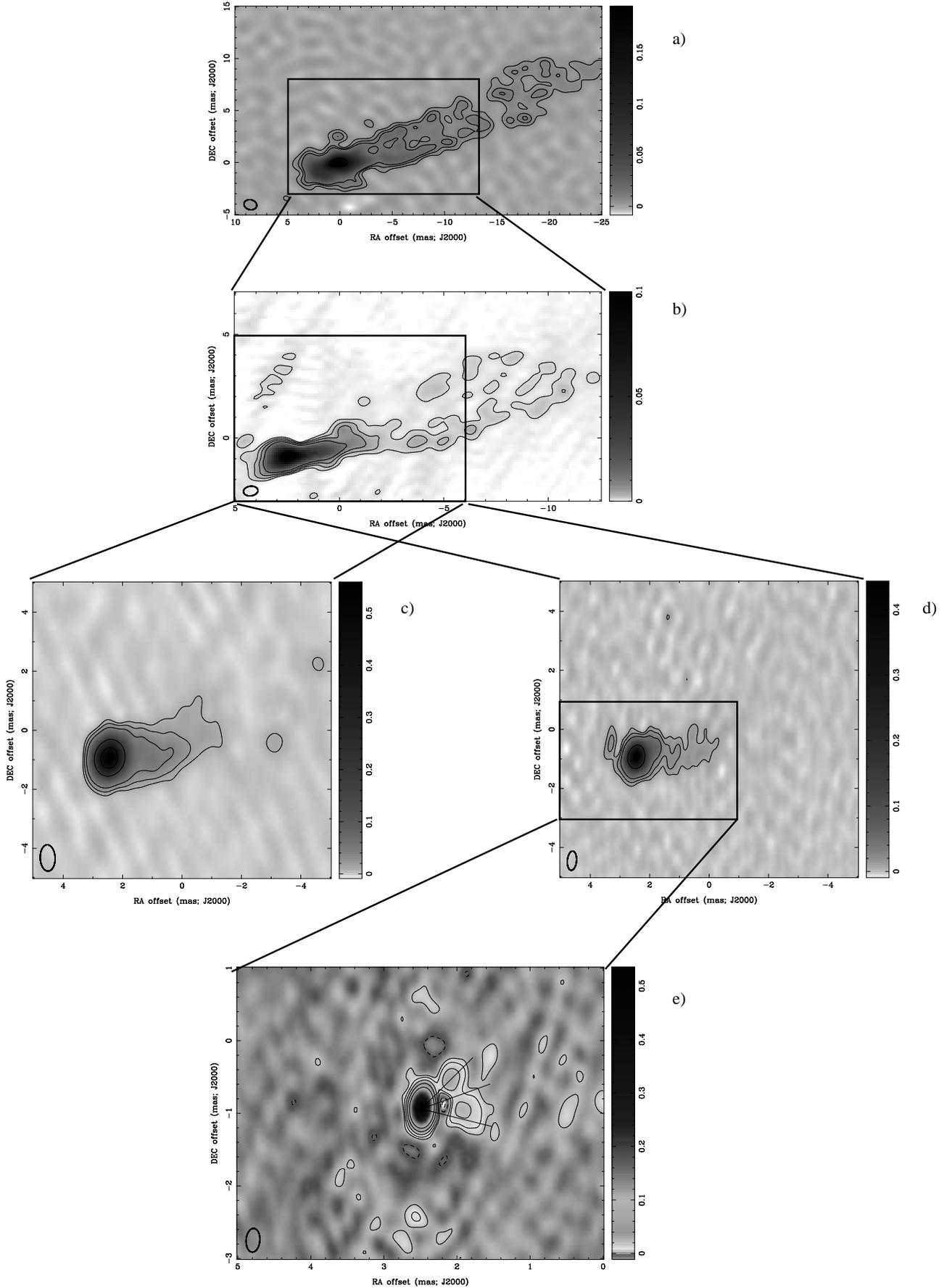,width=0.95\textwidth,angle=0}
%      \psfig{file=fig1.sw.c.eps,width=0.95\textwidth,angle=0}
%  \end{minipage}
\end{center}
\caption{Images of M87 at 
  (a) 1.6\,GHz with a resolution of $1.2\times1.0$
and contours at 5, 10, 15, 150 and 170 mJy/beam, 
  (b) 4.8\,GHz with a resolution of $0.7\times0.5$
and contours at 1, 2.5, 5, 10 and 100 mJy/beam, 
  (c) 15\,GHz with a resolution of $0.9\times0.5$ 
and contours at 10, 20, 40, 200 and 400 mJy/beam,
  (d) 22\,GHz with a resolution of $0.6\times0.3$
and contours at 10, 20, 40, 200 and 400 mJy/beam,  and 
  (e) 43\,GHz with a resolution of $0.3\times0.2$,
and contours at -5, 5, 10, 20, 40 mJy/beam.
  The plots are aligned to the self-calibration centre of the 1.6\,GHz
  image. The beam sizes are shown in the lower left.}
\label{fig:im}
\end{figure}

We present images of M87 at 1.6, 4.8, 15, 22 and 43\,GHz at their
maximum resolutions in figure~\ref{fig:im}.  The full-resolution VSOP
image at 1.6\,GHz has a resolution of 1.23$\times$0.96\,mas at a
position angle (PA) of 77$^\circ$. It was made with uniform weight
binning and no amplitude error weighting to achieve maximum resolution
at the expense of signal to noise. The peak brightness is 0.20\,Jy/beam and
the RMS is 1.1\,mJy/beam.
The low-level emission surrounding the core perpendicular to the jet
axis in the 1.6\,GHz image has not been reported previously. It is not
seen in higher frequency images, consistent with this weak extended
emission having a steep spectral index. This emission would not have
been visible in lower resolution images such as those in \citet{rei89}
and \citet{rei98}.

The 4.8\,GHz VSOP image has a resolution of $0.7\times0.5$ mas at a PA
of $-82^\circ$. It was made with uniform weighting and an amplitude
error weighting of $-1$. The peak brightness is 0.14\,Jy/beam with an RMS of
0.2\,mJy/beam. The image is similar to the VSOP images in
\citet{jun00} and \citet{bir02}.  For analysis of spectral variations it is
necessary to convolve all images with the same beam-size: At 4.8\,GHz
the smoothed image has a resolution which matches the 1.6\,GHz image,
a peak of 0.14\,Jy/beam and an RMS of 0.3\,mJy/beam.

The 15\,GHz VLBA image has a resolution of $0.9\times0.5$ mas at a PA
of $2^\circ$. It was made with uniform weighting and an amplitude
error weighting of $-$1. The peak brightness is 0.56\,Jy/beam with an RMS of
2\,mJy/beam. The smoothed image has a resolution which matches the
1.6\,GHz image, a peak of 0.83\,Jy/beam and a RMS of 2\,mJy/beam.  The
image is similar to those made as part of the VLBA 2\,cm survey
\citep{kel04}.

The 22\,GHz VLBA image has a resolution of $0.6\times0.3$ mas at a PA
of $-$1$^\circ$. It was made with uniform weighting and an amplitude
error weighting of $-$1. The peak brightness is 0.44\,Jy/beam with an RMS of
1.5\,mJy/beam. The smoothed image has a resolution which matches the
1.6\,GHz image, a peak of 0.86\,Jy/beam and a RMS of 3\,mJy/beam.
This can be compared with the 22\,GHz image of \citet{jun95}. The weak
feature about 1~mas to the North-East of the core could belong to the
counter-jet
%(see, e.g., \cite{ly04b}) 
and matches the direction (but not distance) of a similar component
seen at 86~GHz (\cite{krich_04}).

Our 43\,GHz image has a resolution of $0.3\times0.2$ mas at a PA
of $-$4$^\circ$. It was made with uniform weighting and an amplitude error
weighting of $-$1. The peak brightness is 0.53\,Jy/beam with an RMS of 
1.5\,mJy/beam. The smoothed image has a resolution which matches the
1.6\,GHz image, a peak of 0.63\,Jy/beam and a RMS of 5\,mJy/beam.
The peak brightness is significantly higher than that reported by
\citet{jun99} for March 1999 but less than that in \citet{ly04a} observed in
October 2001.  The structure is consistent with those of
\citet{jun99,bir02,ly04a}, although with a lower signal to noise.
Overlaid on our image (figure \ref{fig:im}e) are the opening angles
identified by Junor et al.\ (1999). 

For each of these high resolution images we have calculated the core
brightness temperature in the source rest frame from the core
integrated flux and deconvolved size and these are reported in
table~2. We have taken the Eastern-most component as the core, which
is also the brightest for all but the 1.6\,GHz image. In the 1.6\,GHz
image the brightness temperature of the peak component is
$11\times10^{10}$~K. These values are all well below the inverse
Compton limit and therefore do not require Doppler boosting.
%consistent with absence of apparent superluminal motion

\section{Jet Component Motions}

As mentioned previously, M87 has been well studied using the VLBI technique
and there are many published images with which we can compare those
presented here.

\begin{figure}
%\sfig{overlay.smooth.eps}
%\sfig{match.eps}
  \begin{center}
    \begin{minipage}[t]{0.45\textwidth}
        \psfig{file=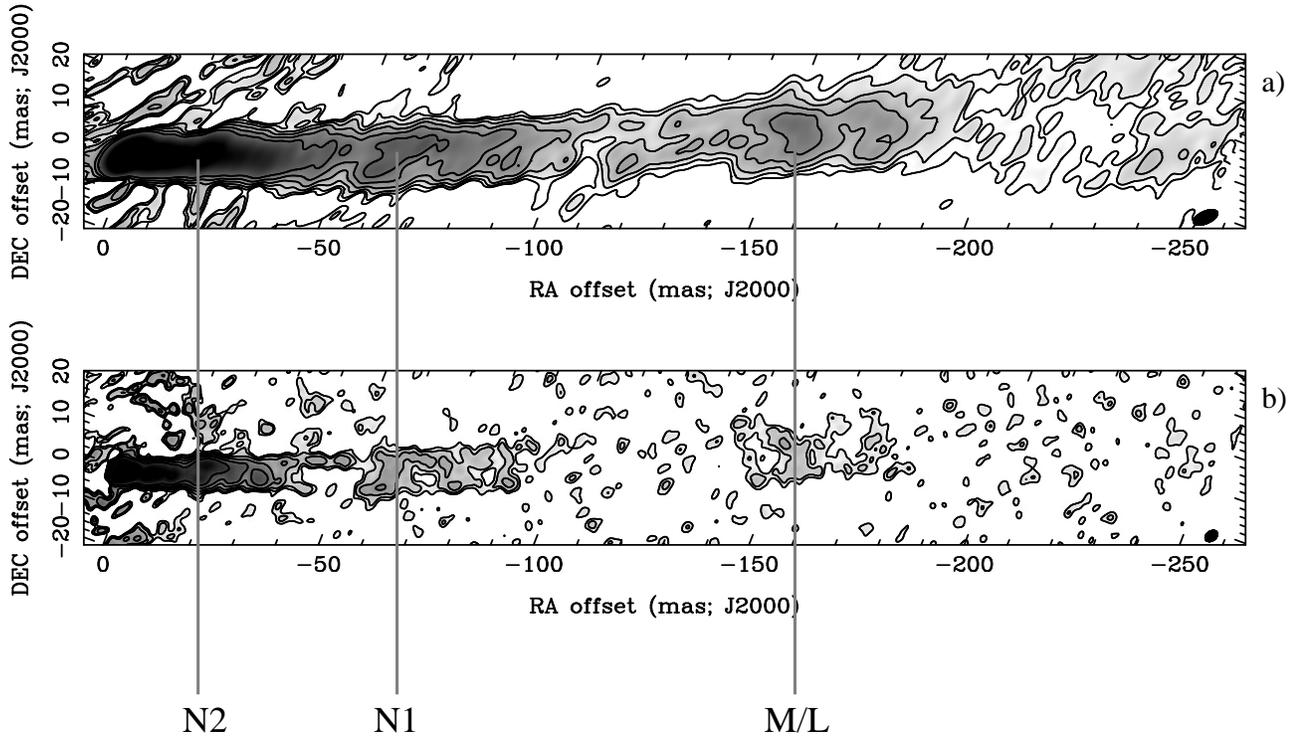,width=17cm,angle=0}
    \end{minipage}
  \end{center}
\caption{M87 at 1.6\,GHz (a) and 4.8\,GHz (b) as observed by the VLBA
  alone and are tapered to 50\% at 50\,M$\lambda$. The 1.6\,GHz
  resolution is $5.9\times2.6$ and the 4.8\,GHz resolution is
  $2.9\times1.9$. The contour levels start at 0.2 and 0.1 mJy/beam,
  respectively, and double with each level. The images have been
  rotated by $-20.62^\circ$ so that the jet runs left to right and and
  the 4.8-GHz image has been aligned so that the peak of the first
  emission along the jet axis coincides. Components N1, N2 and M/L are
  marked. The beam sizes are shown in the lower right.}
\label{fig:match}
\end{figure}

First, we have re-imaged the 1.6\,GHz data with only the VLBA
telescopes, using a Gaussian ({\it u, v}) taper falling to 0.5 at
50\,M$\lambda$.  The resulting image, shown in
figure~\ref{fig:match}a, can be directly compared with the 1.6\,GHz
image of \citet{rei89} from epoch 1984.26, and also with those of
\citet{bj95} at epochs 1988.41 and 1992.47.  The similarities between
the images are startling.  Three jet features were identified by
\citet{rei89}: N2, at a core distance of $\sim$20\,mas, N1 at
$\sim$65\,mas, and M at $\sim$170\,mas.  Component M was subsequently
labeled L in \citet{bj95} and \citet{bir99}. All three components are
visible in figure~\ref{fig:match}, and appear to have remained
stationary over the 15.96 years spanned by these observations (with N1
and N2 having been first identified in lower sensitivity observations
several years earlier). We place the positions of knots N2 and N1 at
19.4 and 64.7\,mas along the jet in the images from \citet{rei89}
(epoch 1984.26), to better than 2\,mas. From our observations we place
N2 and N1 at 18.4 and 62.9\,mas in epoch 2000.22. We can therefore
confirm the upper limits of $-0.02\pm 0.03 c$ and $-0.03 \pm 0.03 c$
for N2 and N1 respectively.

\citet{bir99} reported a motion for knot L of
2.48$\pm$1.06\,mas\,yr$^{-1}$ based on HST observations. This would
suggest motion of at least 22.7\,mas over the 15.96 years, however the
peak of the radio emission in knot L shows no evidence of having moved
at all. In addition, the ``gap'' or minimum in the jet emission
$\sim$110\,mas from the core does not appear to have moved.

Closer to the core, we can compare our VSOP 4.8\,GHz image with the
VSOP image at epoch 1997.97 (\cite{jun00,bir02}).  In addition to N2,
enhancements in the jet emission can be clearly seen at $\sim$13\,mas
and $\sim$6\,mas, the latter corresponding to component B of
\citet{kel04}. These have displayed no discernible motion over the
intervening 2.25 years, although the speed reported for component~B by
\citet{kel04} of 0.14$\pm$0.07\,mas\,yr$^{-1}$ cannot be ruled out.

These observations strongly support the theory that these components
in the radio jet mark the sites of standing shocks.  Filamentary
features, limb-brightening, and side-to-side oscillations have been
reported on a number of occasions (e.g., \cite{owe89,rei89,bir02})
which, as these authors have pointed out, is consistent with a fast
underlying flow. The visible radio emission arises from a less
strongly beamed interaction between the inner jet and the surrounding
sheath. The emission from the sheath must still be Doppler boosted to
some extent as otherwise the counter-jet would be expected to be
easily detectable (e.g., \cite{jun95}).
It is possible to estimate the required Doppler factor from the ratio
of the jet brightness to the noise level.
Following the analysis of \citet{jon96} %,,ryl67, tin98
we use the following expression for smooth continuous jet sheath flow:

\[
{\rm R}=\left( \frac{1+\beta\ {\rm cos}\, \theta}
                    {1-\beta\ {\rm cos}\, \theta} \right)^{2+\alpha}
\]

\noindent
where R is the flux ratio between the jet and counter-jet (the
$3\sigma$ upper limit in this case), $\beta$ is the speed of the jet sheath
in terms of the speed of light
and $\theta$ is the jet angle to the line of sight. 
The spectral index,
$\alpha$, has the opposite sign in this paper than in the reference,
as discussed in section \ref{sec:spec_index}. 
At the peak of the 1.6 GHz emission the brightness is 0.2\,Jy/beam, $\alpha$
is $1.9\pm0.2$ and the image noise is 1.1\,mJy/beam. For $\theta$ we
use the value of $19_{-10}^{+20 \circ}$ of \citet{bir99}. We find R
$>$ 60, therefore the sheath has $\beta > 0.5$. %^{+0.1}_{-0.02}$.

The doppler factor,

\[
\delta = [(1 - \beta^2)^{-1/2} (1 - \beta\ {\rm cos}\, \theta)]^{-1}
\]

%is calculated from the same variables.
%
%
implied by these values is therefore $\delta > 1.6$. 
%Combining this with the highest apparent $\beta$ of 6$c$
%\citep{bir99} requires that $\theta < 17^\circ$.
%%which in turn, assuming a b_app=6 that theta<17. I.e. in agreement
%%with the bir99 value. b_app=2 requires theta<35. 

The discrepancy between 
%the value for $\beta$ we find here and also the
HST-based speed for knot L with the stationary radio features may
perhaps be related to the phenomena observed in the HST observations
for the component HST-1 East, 870\,mas from the core. This knot, which
has a speed of 0.84\,$c$, appears to launch new optical features which
move at speeds of up to $\sim 6 c$ \citep{bir99}. Possibly the
stationary radio feature 170\,mas from the core is the site from which
optical components like knot L are launched (with, in this case,
sub-luminal speeds). We note that there are no published images of M87
at 1.6\,GHz from the period of the HST observations (1994.59 to
1998.55) to determine whether the HST feature has a radio counterpart.

\citet{bir99} suggest that the $\sim 6 c$ motions reflect the
underlying speed of the jet. It is clear that features in the radio
jet between $\sim$7 and 160\,mas appear stationary, in contrast to the
lower limits of 1\,$c$ to 1.6\,$c$ motions recently reported from
observations at 43\,GHz (\cite{ly04b}).  It will be interesting to see
whether the actual speeds close to core are comparable to the
underlying jet speed inferred from the HST observations.
 
Finally, as described in more detail in the next section, we detect a
component $\sim$2\,mas from the core in the 1.6, 4.8, 15 and 22\,GHz
images.  Precise measurement of the core distance is complicated by
frequency dependent opacity effects and, potentially, by the time
intervals between observations, if the lower limits to motions derived
by Ly et al.\ (2004b) still apply at these core distances. We note
that this core distance is similar to that of the component~C of
Kellermann et al.\ (2004).  As noted earlier, component~C was detected
between 1 and 2\,mas from the core by Kellermann et al., but showed no
net motion over the six years. It has been suggested that
``stationary'' components may be standing recollimation shocks, which
could appear to move back and forth to some degree as other jet
components move through them (see, e.g., \citet{jor01} and references
therein): it would be interesting to monitor the movements of this
feature to determine whether there is any relationship with the
motions of the moving jet components observed closer to the core at
43\,GHz.

\section{Image Alignment}

Self-calibrating VLBI data loses the absolute position information,
and produces images centred on the strongest emission, so re-alignment
is required when comparing images at different frequencies.  There are
various approaches to re-aligning observations: we chose the simplest,
which also produces essentially identical answers to more complex
methods.

\begin{figure}
%\sfig{cgslice.shift.zoom.eps}
\sfig{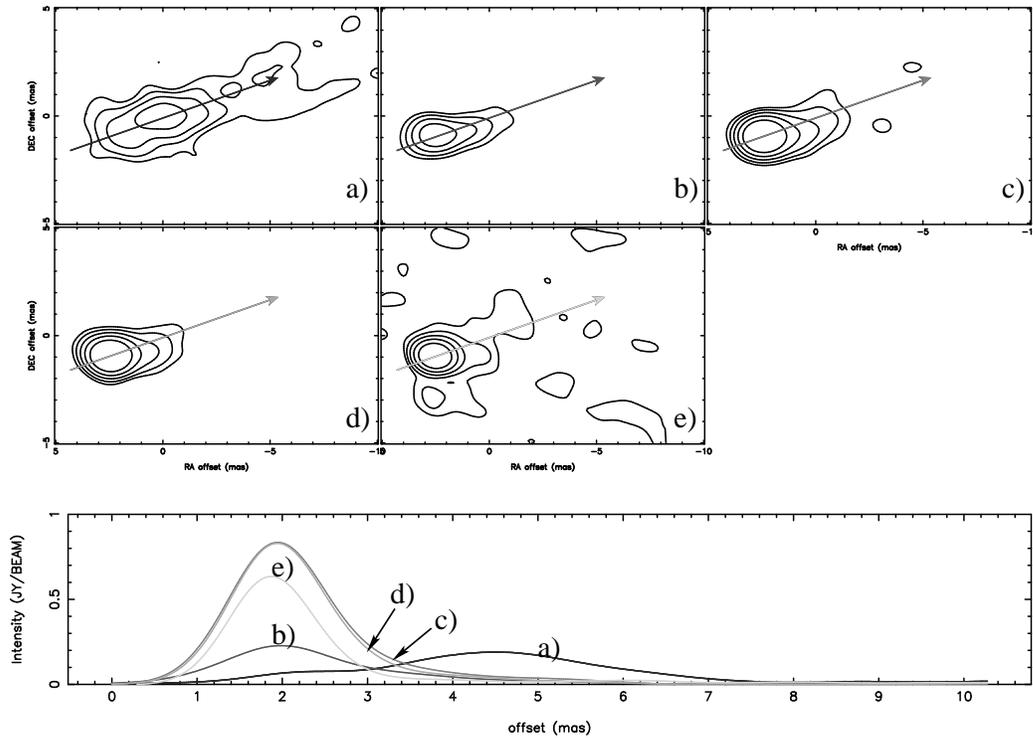}
\caption{Slices through the jets at five frequencies after smoothing
  and alignment. The contour levels start at 15 mJy/beam and double
  with each level. The four images at (b) 4.8\,GHz (c) 15\,GHz 
  (d) 22\,GHz and (e) 43\,GHz have been shifted so that the 
  peak of the first emission
  along the jet axis coincide with that of the 1.6\,GHz (a). }
\label{fig:slice}
\end{figure}

\begin{figure}
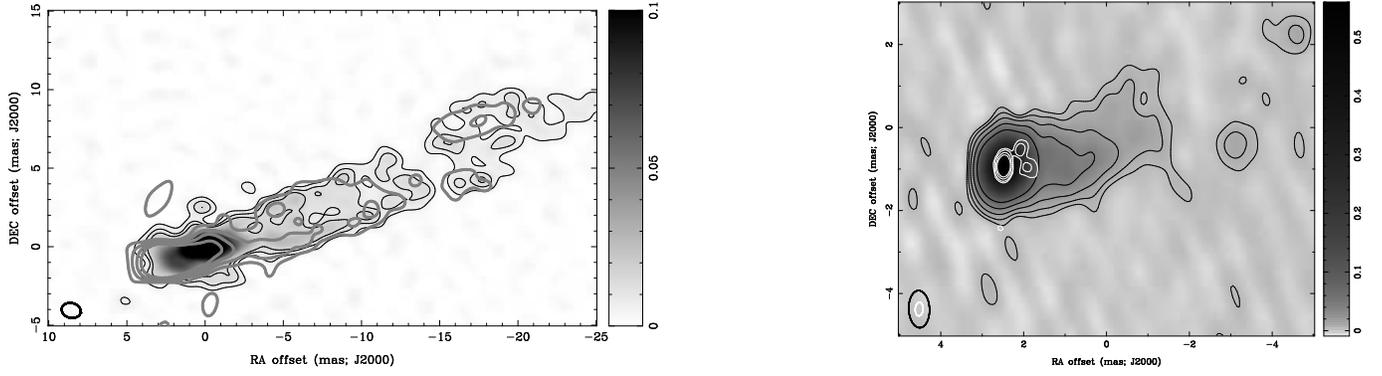

%\dfigs{overlay.l-c.ps}{overlay.15-43.ps}
\dfigs{fig4a.ps}{fig4b.ps}
\caption{Overlays of the shifted images for the frequencies, (a) 1.6\,GHz
  (greyscale and contours) and 4.8\,GHz (contours) at $1.2\times1.0$\,mas,
  and (b) 15\,GHz (greyscale and contours) and 43\,GHz (white contours) at
  $0.9\times0.5$ and $0.3\times0.2$ mas. The contours are 5, 10 and 15
  mJy/beam for 1.6\,GHz; 2, 4, 8 and 16 mJy/beam for 4.8\,GHz; 5, 10,
  20, 40 and 80 mJy/beam for 15\,GHz; 10, 20, 50 and 100 mJy/beam for
  43\,GHz. The beam sizes are shown in the lower left.}
\label{fig:ol}
\end{figure}

The jet in M87 is straight, so does not allow the alignment on a
frequency-independent feature such as a bend.  The region closest to
the core could be approximately decomposed into two components between
1.6 and 22-GHz, however, so it was possible to identify and align on
the core and innermost jet component.  The latter, which is
$\sim$2\,mas from the core, is the brightest feature at 1.6\,GHz, with
the core being brighter at 4.8\,GHz and above. Because of the poor
representation, and therefore positional accuracy, of the Gaussian
components (particularly for the 1.6\,GHz image) we abandoned
unconstrained image alignment. We finalised on identifying 1-D
Gaussian components along the jet axis. This allowed us to produce
images aligned on the maximum of the core, i.e., the Eastern-most
image component, along the jet axis. Slices through the core and along
the inner jet of the aligned images are plotted in
figure~\ref{fig:slice} and overlays of the four lowest frequencies are
overlaid in figure~\ref{fig:ol}.

Core shifts are ignored in this approach, but these are expected to be
small (\cite{lob98a}). Lobanov (2005, personal communication) predicts
a core shift of 0.015\,mas between 5 and 22\,GHz for M87 assuming
nominal jet parameters. The offset between 1.6\,GHz and 43\,GHz is
0.1\,mas, in line with our errors.  In support of our approach, we
note the good alignment between the edges of the jet in the
first 10~mas in the 1.6 and 4.8\,GHz images.  There is a broad
component $\sim$ 6\,mas
from the 1.6\,GHz core (described in the previous section)
%($-$5, 2.6) mas 
which aligns between the 1.6 to 22\,GHz images, but it cannot be
identified in the 43\,GHz image. The 15, 22 and 43\,GHz images are
very similar and have only a small relative shift.

It is worth noting that shifting the core site away from the peak of
the 1.6\,GHz emission does not alter the calculations of the opening
angle, such as those described in \cite{rei89}.  An opening angle of
6.9~degrees was derived by Reid et al.\ from the increase of the jet
% comments
%ou give an opening angle of 6.9 degrees. If this
%corresponds to a Mach cone, the Mach number
%would be M= 1/sin(phi) = 8.3. 
width along the jet axis, and a core shift such as that described here
therefore results in a change in the fitted core diameter at the
origin from 1.7\,mas to 1.2\,mas. This core size matches closely the
predicted size resulting from interstellar and interplanetary
scattering of $\approx$1\,mas \citep{coh_74}.

\subsection{The spectral index}
\label{sec:spec_index}

\begin{figure}
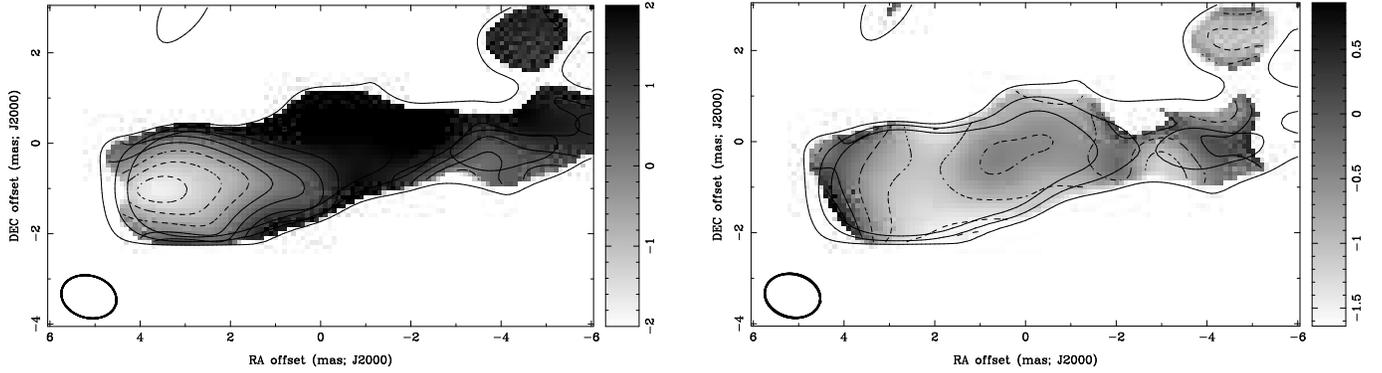

%\dfigs{si.l-c.ps}{si.5-15.ps}
\dfigs{fig5a.ps}{fig5b.ps}
\caption{Spectral index between (a) 1.6 and 4.8\,GHz, and 
  (b) 4.8 and 15\,GHz. Images have been smoothed to $1.2\times1.0$ mas, 
  with a position angle of 81$^\circ$. The outline of the beam is in the
  bottom left of the image. Contours at every 0.5 of the spectral
  index are overlaid. Three 4.8\,GHz contours also have been plotted
  at 2.5, 5 and 10 mJy/beam.}
\label{fig:si}
\end{figure}

The spectral index, $\alpha$, between 1.6 and 4.8 GHz and 4.8 to
15\,GHz, after smoothing to a common resolution are presented in
figure~\ref{fig:si}, defined in the sense $S \propto
\nu^{-\alpha}$. The extended emission around the 1.6\,GHz image,
mentioned in Section 4, and not visible at higher frequencies, implies
a spectral index of greater than 2 to the North and South of the
region mapped.
Between the higher frequencies we have a negative spectral index
across most of the source ($-0.6\pm0.4$), as would expected given that
M87 is brighter at 15~GHz than 4.8~GHz (see
figure~\ref{fig:slice}). Between 1.6 and 4.8~GHz the spectral index is
more complex, with a inverted component ($-1.1\pm0.4$) closest to the
core and a steep component ($2.1\pm0.2$) near the peak of the low
frequency emission.
This emphasizes the differences in character of these two emission
regions, and the analytical importance of the lower frequency, high
resolution VSOP observations.

\begin{figure}
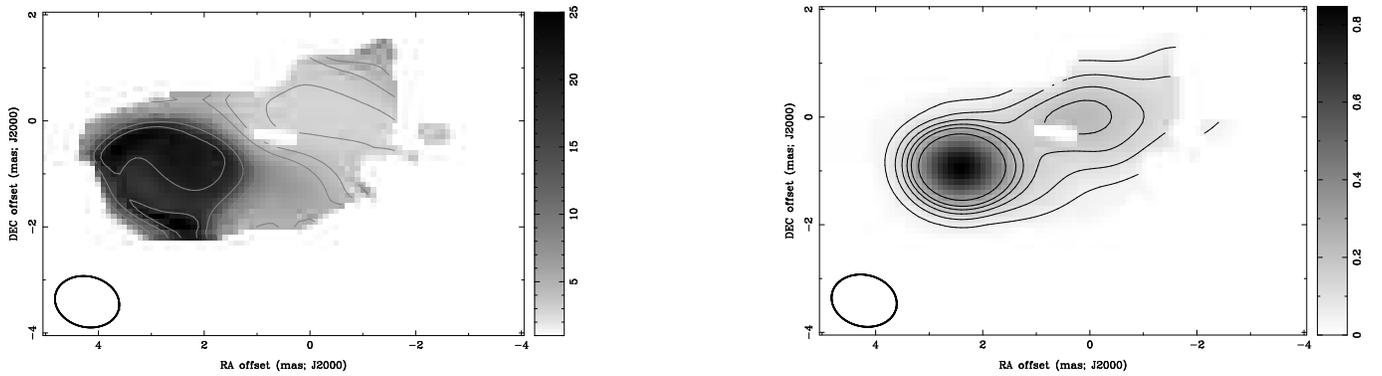

%\dfigs{m87.vmax.ps}{m87.smax.ps}
\dfigs{fig6a.ps}{fig6b.ps}
\caption{a) Turnover frequency calculated from the pentachromatic data with
  contours at 3, 4, 5, 10, 15, and 20\,GHz and b) brightness at
  that frequency with contours at 5, 10, 15, 20, 25 and 35\,mJy/beam.
  for M87 based on fit to log($S$), log($\nu$). A lower cutoff of 0.1 GHz and
  a high cutoff of 1000 GHz was used in the fitting.
%  Contours for the
%  1.6 and 22 GHz emission are overlaid. This high lights the two
%  peaks. One, closer to the AGN, dominant at 5 GHz and above, the
%  other dominant at 1.6 GHz. 
}
\label{fig:vmax}
\end{figure}

For the spectral fitting we used our pentachromatic dataset across the
AGN core, and limits for the cutoff frequencies. Because there are
great difficulties is measuring these cutoffs we have fitted them with
very large errors, but forcing the spectral fits to have positive
curvature.
We have followed the methods of \citet{lob98b} to calculate the
turnover frequencies as the jet develops along the axis. We fitted
second order polynomials to the log($S$) log($\nu$) data, with assumed
spectral cutoffs of 0.1 and 1000\,GHz, as shown in figure \ref{fig:vmax}, 
which shows the derived maximum frequency, $\nu_{max}$ and the flux at that
frequency, $S_{max}$.
A range of cutoff frequencies were trialled before settling on these
values.  Significantly higher values for the upper, or smaller values
for lower, cutoff distorted the spectrum sufficiently to discount them
as possibilities.  The two centres of emission clearly indicated in
\ref{fig:vmax}b, and suggested by the two-component image fits, place
the low frequency peak within the collimation zone described by Junor
\etal (1999).
We have estimated the systematic effects of the errors in image
alignment by performing multiple fits with introduced random shifts of
$\pm0.1$~mas in the alignment and comparing these to the plotted
fits. We find the errors are the order of 20\% for both $\nu_{max}$
and $S_{max}$.

\begin{figure}
\begin{center}
%\sfig{B_1D.ps}
  \begin{minipage}[t]{1.0\textheight}
      \psfig{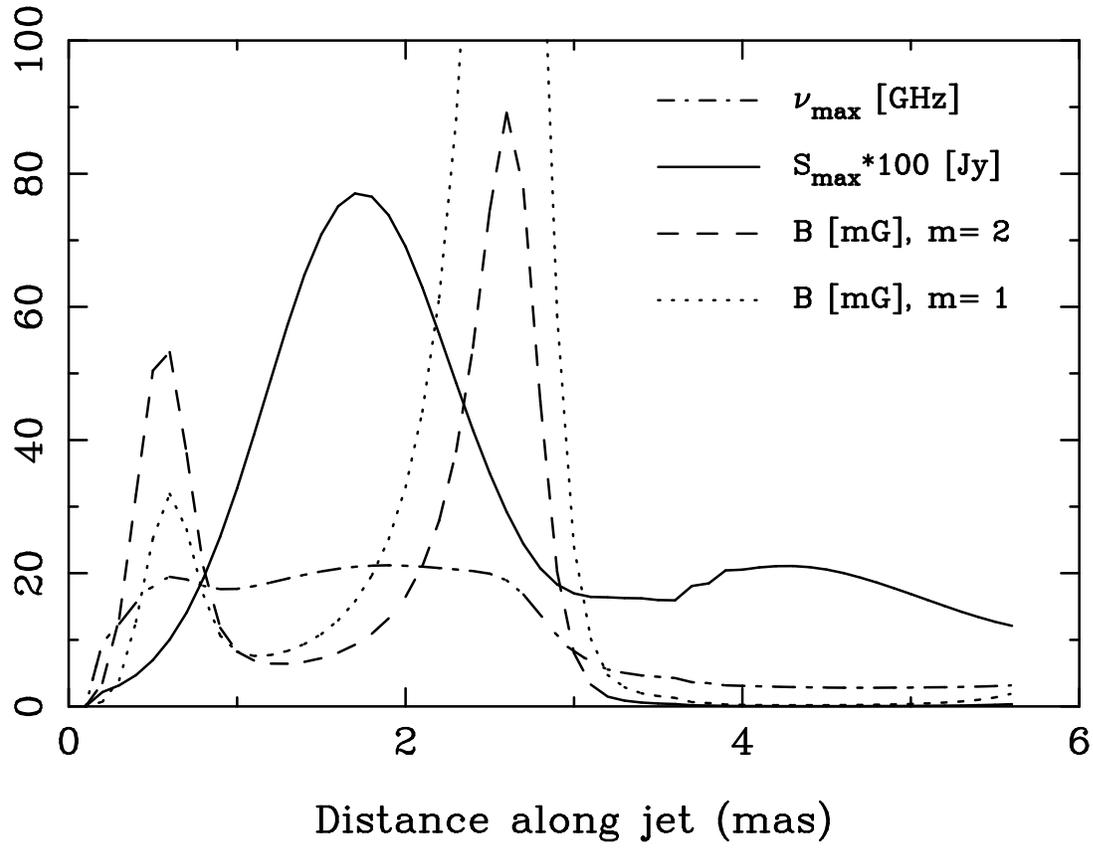}

%height=0.75\textheight
  \end{minipage}
\end{center}
\caption{The one dimensional average across the jet centre of
  $\nu_{max}$, the turnover frequency in GHz, and $S_{max}$, the brightness at
  that frequency in Jy/beam (scaled up by a factor of 100) and the
  magnetic field in mG, assuming an underlying ambient field of
  $B_{amb} \propto r^{-2}$, or $\propto r^{-1}$.}
\label{fig:B}
\end{figure}

Following the shocked emission approach outlined in \citet{caw_91} and
\citet{lob98b} (equations 17 and 18), we can relate the turn-over flux
density and frequency, together with the distance down the jet, to the
magnetic field strength:

\[ B(r) = C_0 \nu_{max}^5 r^{4-m} S_{max}^{-2} \]

\noindent
where $C_0$ is a constant of proportionally, $\nu_{max}$ the turn-over
frequency, $S_{max}$ the flux at this maximum, $r$ the distance down the
jet and $m$ the power of the background ambient magnetic field. 
We use the values for the core together with the lower limit to the
average magnetic field strength for the core of 22\,mG derived by
\citet{bir91} to determine $C_0$.  We averaged the $\nu_{max}$ and
$S_{max}$ across the jet and plot these in figure \ref{fig:B}. The
figure also plots the derived magnetic from these two values assuming
that the ambient background magnetic field is proportional to
$r^{-m}$, where, following \citet{lob98b} we consider cases with both
$m=2$ and $m=1$.  The uncertainties in $\nu_{max}$ and $S_{max}$,
and the high sensitivity of $B$ to these, lead us to be cautious about
the derived quantities. However, as a qualitative guide to the field
required to generate the observed emission, the results are robust and
repeatable even in our randomised datasets.
Two peaks in the magnetic field are seen, whether ambient field varies
as the square or linearly. The first peak occurs just before the high
frequency peak of emission and the second occurs just before the low
frequency peak. The magnetic field between these two peaks is
$\sim$10\,mG. This leads us to the speculation that core is
delimited by magnetic field compression at the boundaries, that the
field is roughly uniform within the core, and that the low frequency
peak is outside this core, and in the collimation zone.
We note that \citet{lob98b} derived a similar behaviour
for the magnetic field of 3C345.

\section{Summary and Discussion}

We have presented images of M87 at 1.6 and 4.8\,GHz from VSOP
observations, and at 15, 22 and 43\,GHz from VLBA observations
three weeks later. The
combination of lower frequency and high resolution provided by the
space baselines allowed us to directly measure the turn-over
frequency, rather than extrapolate it.
The spectral index maps show distinct differences between the core and
the component $\sim$2\,mas from the core. We have used the five
frequencies to fit the spectrum of the source, pixel by pixel. From
this we derive the frequency of maximum emission and the flux at that
frequency. We have combine these values to calculate the variation of
the magnetic field strength across the core and into the jet. Taking a
reasonable set of assumptions, the magnetic field peaks at the outer
edges of the core, and is flat across it.

Applying a ($u,v$) taper to the 1.6\,GHz data and restoring with a
larger beam resulted in the detection of jet components up to 160\,mas
from the core, revealing that these features have appeared stationary
over the last 16 years.  Comparison of the 4.8\,GHz image with the
earliest VSOP image of M87 indicates that components closer to the
core have also appeared stationary on a 2.25 year timescale.

The reported detection of TeV gamma-rays from M87 invites comparison
of these jet speeds with the sub-luminal and stationary components
observed in other AGN detected at TeV energies
\citep{edw02,pin04}. The comparison is complicated by the fact that
M87 is an intrinsically much weaker TeV source \citep{aha03}.
\citet{bai_01} consider the contrasting effects of M87's proximity but
larger jet angle to the line of sight, $\theta$. They note that if the
bright TeV sources Mkn\,421 or Mkn\,501, both at distances of
$\sim$150\,Mpc and for which they adopt $\theta \lesssim 6^\circ$,
were at the distance of M87 but viewed at $\theta = 35^\circ$, they
would be weak but detectable. \citet{bir99} however, derive $<
19^\circ$ based on the HST observations of apparent speeds of 6\,$c$,
which would increase the predicted flux (keeping other assumptions
constant) by a factor of 20.  Given all the uncertainties, it is at
least plausible that M87 is a detectable (if variable; \cite{leb04})
TeV source.

Assuming this is the case, we can then examine the explanations
proposed for the low jet motions in TeV gamma-ray sources
\citep{edw02,pin04} for the specific case of M87.  M87 shows evidence
for a significant jet opening angle \citep{rei89}, which \citet{gop04}
have pointed out can result in apparent speeds much slower than the
actual speed due to variation of boosting across the width of the
jet. However, the effectively stationary jet motions observed on the
mas-scale would suggest this model does not apply in this case. There
would appear to be evidence for decelerating jet model of
\citet{geo03} in the recent reported jet speeds close to the core at
43\,GHz, however one must then invoke reacceleration at large core
distances to explain the superluminal motions detected in HST
observations. The most natural explanation for M87 would appear to be,
as discussed earlier, an invisible jet spine with the observed jet
components being standing shocks in the sheath of surrounding
material.

The apparent component motions observed at 43\,GHz thus offer the
promise of studying collimation region and, possibly, the transition
region from significant motions to apparently stationary features in
the parsec-scale jet. The planned VSOP-2 mission (\cite{hir04}) will
enable observations at 43\,GHz with even higher resolution, which will
enable the acceleration and collimation of this important radio jet to
be examined even more closely.

%\acknowledgments

\vskip6mm

We gratefully acknowledge the VSOP Project, which is led by the
Institute of Space and Astronautical Science of the Japan Aerospace
Exploration Agency, in cooperation with many organizations and radio
telescopes around the world.  RGD acknowledges support from the Japan
Society for the Promotion of Science, and the hosting by Observatorio
Astron\'omico Nacional, Espa{\~n}a, during the preparation of this
paper. The NRAO is a facility of the National Science Foundation,
operated under cooperative agreement by Associated Universities, Inc.

\onecolumn

%% \begin{figure}
%% \sfig{cluster_centre.eps}
%% \caption{Cluster centre after shifting to match the first light}
%% \label{fig:gf}
%% \end{figure}
%% \begin{table}
%% \begin{center}
%% \begin{tabular}{lccc}
%% \hline
%% frequency (GHz) & shift (mas) (first light) & (1D Gaussian fit)&2D fit\\
%% 1.6 &     0,0    &    0,0    &      0,0    \\
%% 4.8 & -1.81,-0.68&-2.33,-0.88&  -1.80,-1.34\\
%% 15  & -2.09,-0.78&-2.41,-0.91&  -2.22,-1.41\\
%% 22  & -2.22,-0.84&-2.40,-0.90&  -2.51,-1.43\\
%% 43  & -2.51,-0.95&-2.49,-0.94&  -3.07,-1.46\\
%% \hline
%% \end{tabular}
%% \caption{Shifts in $\alpha_{RA}$ and $\delta$ to align on xxxx.}
%% \end{center}
%% \end{table}

\end{document}